\documentclass[10pt, letterpaper, twocolumn]{article}
\usepackage{ol2}
\usepackage{amsmath}
\usepackage{graphicx}
\usepackage{color}

\begin{document}

\twocolumn[
\title{Broadband squeezing of quantum noise in a Michelson interferometer with Twin-Signal-Recycling}

\author{Andr\'{e} Th\"uring, Christian Gr\"af, Henning Vahlbruch, Moritz Mehmet, Karsten Danzmann, and Roman Schnabel}

\address{Institut f\"ur Gravitationsphysik, Leibniz Universtit\"at Hannover and Max-Planck-Institut f\"ur Gravitationsphysik (Albert-Einstein Institut), Callinstra\ss e 38, 30167 Hannover, Germany}

\begin{abstract}
Twin-Signal-Recycling (TSR) builds on the resonance doublet of two optically coupled cavities and efficiently enhances the sensitivity of an interferometer at a dedicated signal frequency. We report on the first experimental realization of a Twin-Signal-Recycling Michelson interferometer and also its broadband enhancement by squeezed light injection. The complete setup was stably locked and a broadband quantum noise reduction of the interferometers shot noise by a factor of up to 4\,dB was demonstrated. The system was characterized by measuring its quantum noise spectra for several tunings of the TSR cavities. We found good agreement between the experimental results and numerical simulations. 
\end{abstract}

]

Advanced laser interferometers use optical resonators in order to improve the signal to quantum noise ratio. All current interferometric gravitational wave detectors LIGO, TAMA, VIRGO, and GEO\,600 
apply the technique of \emph{power-recycling} (PR) which uses a mirror between the laser source and the interferometer's central beam splitter in order  to resonantly enhance the circulating laser power. The GEO\,600 detector topology also incorporates an additional mirror in the detection port, which retro-reflects the \emph{signal} fields back into the interferometer. The microscopic position of this so-called signal-recycling (SR) mirror determines what signal frequencies get resonantly enhanced. For an overview we refer to \cite{Aufmuth05}.

A modulation signal at the frequency $\omega$ produces an upper sideband field at $\omega_0+\omega$ and a lower sideband field at $\omega_0-\omega$ where $\omega_0$ is the light frequency of the laser source. Thus, if the recycling frequency is higher than the bandwidth of the SR cavity, only the upper \emph{or} the lower sideband field can be enhanced whereas the counterpart is suppressed, thereby loosing half of the maximum signal. A completely independent approach to improve the signal to quantum noise ratio of a laser interferometer is the injection of squeezed states of light into the dark signal port of the interferometer \cite{CavesPRD23,XiaoPRL59,GrangierPRL59}. It was demonstrated that this technique is compatible with power-recycling~\cite{McKenziePRL88}. However, for the case of single sideband SR, it turns out that the squeezing angle is rotated by the SR cavity and no broadband quantum noise reduction is achievable with injected fields of an ordinary white squeezing spectrum. Harms \textit{et al.}~\cite{HarmsPRD68} showed that this problem can be solved by adding a Fabry-Perot filter cavity to the interferometer set-up with cavity parameters identical to the SR cavity. The broadband reduction of shot noise in a Michelson interferometer with single sideband SR plus an additional filter cavity was then successfully demonstrated in~\cite{VahlbruchPRL95}. Recently, we have theoretically shown that an interferometer topology exists that, firstly, is able to resonantly enhance both, the upper and lower sidebands of a dedicated signal frequency band, and secondly does not require an additional filter cavity for a broadband squeezing enhancement~\cite{ThuringOL32}. This topology builds on the resonance doublet of two optically coupled TSR cavities and was therefore termed \emph{Twin-Signal-Recycling} (TSR).\\ 
In this Letter we report on the first experimental realization of a squeezed light enhanced TSR interferometer. 

\begin{figure}
\centerline{
\includegraphics[scale=0.55]{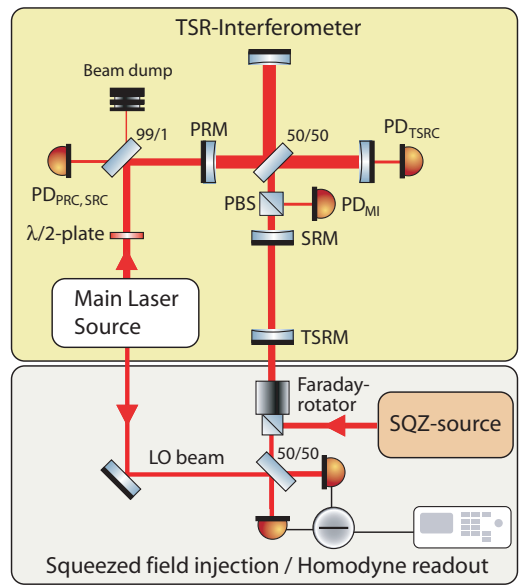}} 
\caption{\label{schema}(Color online) Schematic of the experimental setup. PD: photo diode; PBS: polarizing beam splitter; PRM, SRM, TSRM: power-, signal-, and twin-signal-recycling mirror. LO: local oscillator. SQZ: Squeezed light.}
\end{figure}

The experimental setup is illustrated in Fig.~\ref{schema}. The main laser source was a Nd:YAG non-planar ring oscillator with a continuous-wave single-mode output power of up to 2\,W at 1064\,nm. The laser beam was transmitted through a ring resonator providing spatial and high frequency amplitude and phase noise filtering. A major part of the laser power was frequency doubled in a non-linear resonator; details can be found in~\cite{ChelkowskiPRA71}. About 1.5\,mW of the filtered beam at the fundamental frequency served as local oscillator beam for homodyne detection.\\
Approximately 35\,mW \textit{s}-polarized light was injected into the TSR Michelson interferometer through the power-recycling mirror (PRM). The PRM had a power reflectance of 90\% and, together with the interferometers end mirrors having a reflectance of 99.92\,\%, formed the carrier resonating power-recycling cavity (PRC). The length of the PRC was approximately 1.21\,m corresponding to a free spectral range of 123.6\,MHz . Its finesse  was measured to 60 leading to a bandwidth of about 2\,MHz when the interferometer was locked to a dark fringe at the signal output port. The PRC length, the interferometer dark port, and the TSR cavities lengths were stabilized by RF-modulation/demodulation schemes and by piezo-actuation of mirror positions. For this purpose the 35\,mW \textit{s}-polarized interferometer input beam was phase modulated at 15\,MHz as well as 125.6\,MHz and accompanied by a 10\,mW \textit{p}-polarized beam of the same frequency carrying the same two phase modulations. The error signal for the PRC length control was gained from the 15\,MHz modulation on the \textit{s}-polarized light reflected from the PR cavity ($\textrm{PD}_\textrm{PRC}$ in Fig.~\ref{schema}). The Michelson interferometer dark fringe error signal was derived from the 125.6\,MHz modulation on the \textit{p}-polarization ($\textrm{PD}_\textrm{MI}$) detected behind a PBS  placed between the interferometer beam splitter and the SRM in order to decouple the polarization modes. The length control of the two coupled TSR cavities built by the end mirrors, the SRM and the TSRM utilized the sidebands at 125.6\,MHz in the \textit{s}-polarization detected in reflexion of the PRM ($\textrm{PD}_\textrm{SRC}$) and in transmission of one end mirror ($\textrm{PD}_\textrm{TSRC}$), respectively. Since the sidebands at 125.6\,MHz were within the bandwidth of the PRC and were thus enhanced, an interferometer arm length difference of 7\,mm was sufficient to obtain proper error signals for the positions of SRM and TSRM. The optical length of the cavity built by SRM and the end mirrors was approximately 1.19\,m, the length of the resonator formed by the SRM and the TSRM was about 1.26\,m. With the SRM power reflectance of 90\% this leads to an expected frequency splitting of approximately 6.1\,MHz. \\
Another 15\,mW of the filtered main laser beam were used for length control of the squeezed light source cavity (not shown in Fig.~\ref{schema}). The squeezed light source used type\,I optical parametric amplification (OPA) and was realized as a single-ended standing-wave nonlinear resonator formed by two mirrors and a PPKTP-crystal, similar to the source described in \cite{ChelkowskiPRA71}. The squeezed light source was pumped by 450\,mW of 532\,nm radiation and provided a dim amplitude squeezed field of approximately 45\,$\mu$W at 1064\,nm. The squeezed field passed a combination of a polarizing beam splitter, a $\lambda$/2-plate and Faraday rotator, and was matched to the interferometer mode with a visibility of greater than 99\%. Since the TSR interferometer was locked at its dark signal port, the squeezed field was reflected by the interferometer and, due to the polarizing optics mentioned before, was found in the interferometer signal output spatial mode~\cite{McKenziePRL88}. This combined mode was matched to the local oscillator beam of the homodyne detector with a measured visibility of about 95\,\%.
\begin{figure}
\centerline{
\includegraphics[scale=1]{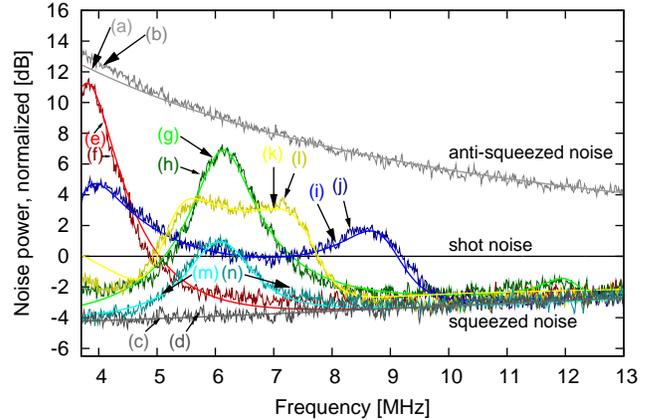}}
\caption{\label{doublets}(Color online) Comparison of measured and simulated quantum noise spectra of the squeezing enhanced TSR interferometer at suboptimal OPs: Simulated and measured \emph{anti-squeezing} spectra (a,b) as well as simulated and measured \emph{squeezing} spectra (c,d) when the TSR interferometer was tuned to anti-resonance.  (e) - (n) Simulated and measured spectra when the TSR cavities were locked close to but not \emph{at} the optimum OP.}
\end{figure}

\begin{figure}
\centerline{
\includegraphics[scale=1]{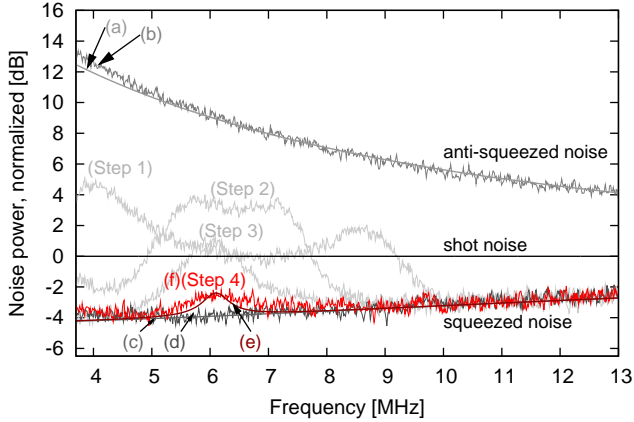}}
\caption{\label{detTSR}(Color online) Measured (f) and simulated (e) broadband squeezing of quantum noise in the TSR interferometer achieved at the optimum OP. (a) - (d) See caption of Fig.~\ref{doublets}}
\end{figure}

Fig.~\ref{doublets} shows quantum noise spectra of the TSR interferometer for different suboptimal tunings of the TSR cavities. Curves (a) to (d) provide upper and lower bound references and represent the noise spectra with strong under-coupling to the TSR cavities. For these measurements these coupled cavities were on anti-resonance. For all the other curves in Fig.~\ref{doublets} the TSR cavities lengths were close to, but not \emph{at} to the optimum operation point (OP) and therefore produced dispersion and a frequency dependent rotation of the squeezing ellipse. The noise spectra for the optimum OP of the TSR cavities are shown in Fig.~\ref{detTSR} providing the lowest quantum noise floor for a strong coupling to the TSR cavities. The optimum OP is realized if the detunings of the two TSR cavities are such that the resonances of upper and lower sidebands are arranged symmetrically around the carrier frequency.  Only in this case no dispersion is produced by the TSR cavities, the squeezing ellipse is not rotated, and a broadband shot noise reduction from frequency independent squeezed light injection is achieved, as proposed in~\cite{ThuringOL32}. Starting with the interferometer stabilized at suboptimal tunings, the optimum OP was reached by smoothly adjusting the offsets of the error signals for SRM and TSRM. As illustrated in Fig.~\ref{detTSR} the resonance peaks were brought together until the rotation into the anti-squeezed quadrature vanished. The remaining bump at approximately 6.1\,MHz arises \emph{not} from quadrature rotation but from losses inside the interferometer. Notice, that the squeezing is just affected by interferometer losses at sideband frequencies fulfilling the resonance condition of the coupled TSR resonators and thus entering the interferometer. Hence from this measurement the frequency splitting can be deduced to approximately 6.1\,MHz which agrees with the TSR parameters. 

The smooth lines in Fig.~\ref{doublets} and Fig.~\ref{detTSR} represent our numerical simulations which are based on the mirror reflectivities, macroscopic cavity lengths, cavity detunings, optical loss and the homodyne angle. The cavity detunings for each measurement were derived from the resonance frequencies in the noise spectra.  The values for the homodyne angle and for the optical loss due to an interferometer dark port offset were then fitted, and were found to match the results of independent measurements within their uncertainties. For the experimental data shown in Fig.~\ref{doublets} the uncertainty for the actual homodyne angles were untypical high, for two reasons. Firstly, the quadrature angle could not be determined by simply minimizing the squeezed noise variance in Fig.~\ref{doublets}  because of the quadrature rotation and frequency dependent quantum noise. Secondly, when varying the TSR cavities detunings the mode degeneracy and therefore the power in residual transversal higher order modes caused an offset change of the homodyne quadrature error signal. For this reason the homodyne angle also varied from trace to trace in Fig.~\ref{doublets}. For the simulations of curves (i) and (k) we used detection angles of 9  and 12.6 degrees, respectively, and detection angles between 0 and 1.4 degrees for all other curves. The optical loss of the interferometer was also dependent on the cavity detunings. Firstly, the error signals for the length control of the two TSR cavities and the PR cavity coupled with the error signal of the dark port control. Notice, that an offset from the perfect dark port condition increases the loss on the squeezed states. Secondly, the changes in the degeneracy with higher order transversal modes changed the strength of the according loss channels. The simulation in Fig.~\ref{doublets} is based on values for the loss inside the TSR interferometer in a range from 0.4 to 3.9\%.\\
In contrast to Fig.~\ref{doublets}, the simulation for the optimum OP shown in Fig.~\ref{detTSR} was not based on any fitting. Here, the homodyne quadrature angle could easily be optimized and precisely locked to zero degrees by minimizing the noise floor. Spatial mode degeneracy was not observed and the dark port condition could be realized by minimizing the loss peak. The remaining loss corresponded to the minimum loss in our setup of 0.4\% which was given by the transmittivities of the end mirrors and non perfect anti-reflection coatings of the cavity mirrors and beam splitters.

In summary, we experimentally demonstrated that an interferometer with Twin-Signal-Recycling (TSR) and properly chosen detunings of its two coupled cavities shows a broadband quantum noise reduction from the injection of non-classical field with an white squeezing spectrum. All longitudinal degrees of freedom were electronically stabilized by modulation/demodulation schemes. The TSR interferometer was analyzed by measuring the quantum noise spectrum for several of its OPs. Our results are in good agreement with numerical simulations. Since the squeezing enhanced TSR interferometer is also able to resonantly enhance both, upper and lower signal sideband fields, it is favorable compared with the corresponding single sideband conventional SR interferometer and might find applications where high, narrow-band sensitivities are required.
 
This work has been supported by the Deutsche Forschungsgemeinschaft and is part of
Sonderforschungsbereich 407. We also acknowledge Albrecht R\"udiger who invented the name \emph{Twin-Signal-Recycling}.


\end{document}